# Enhancing the Figure of Merit in Te-doped FeSb$_2$ through nanostructuring

M. Pokharel,[1] H. Z. Zhao,[2] M. Koirala,[2] Z. F. Ren,[2*] and C. Opeil[1*]

[1]Department of Physics, Boston College, Chestnut Hill MA 02467-3800

[2]Department of Physics and TcSUH, University of Houston, Houston, TX 77204-5005

*To whom correspondence should be addressed

## Abstract

We study the thermoelectric properties of Te-doped FeSb$_2$ nanostructured samples. Four samples of stoichiometry FeSb$_{1.84}$Te$_{0.16}$ were prepared by a hot press method at temperatures of 200, 400, 500, and 600 °C. Te-doping enhances the dimensionless figure of merit (*ZT*) on FeSb$_2$ via two mechanisms. First, a semiconductor to metal transition is induced, which enhances the value of the power factor at low-temperatures. Second, the thermal conductivity, which was already reduced in nanostructured FeSb$_2$ samples, is further reduced by increased point defect scattering through the n type substitution of Sb site by Te atom. The combined effect results in a *ZT* = 0.022 at 100 K, an increase of 62% over the *ZT* value for the optimized Te-doped single crystal sample. Hall coefficient and electrical resistivity measurements reveal a decreased mobility and increased concentration of the carriers in the doped sample.

## Introduction

Due to its unusual magnetic and electronic transport properties, the narrow-gap semiconductor FeSb$_2$ has been extensively studied in the past few decades [1-3]. In the recent years, FeSb$_2$ has attracted considerable attention as a thermoelectric material. Because of its unusually large Seebeck coefficient of ~ 45,000 μVK$^{-1}$ at ~ 10 K [4], FeSb$_2$ is now considered a potential candidate for thermoelectric cooling applications at cryogenic temperatures. The coefficient of performance of a Peltier- cooler increases with the increase in the figure of merit (*ZT*) of the material. $ZT = S^2\rho^{-1}\kappa^{-1}T$, where *S* is the Seebeck coefficient, ρ is the electrical resistivity, κ is the thermal conductivity, and *T* is the absolute temperature. The *ZT* values for single crystal FeSb$_2$ are rather low due to the large thermal conductivity. For undoped FeSb$_2$ single crystals, the reported *ZT* values are around 0.005 at ~ 10 K [4]. Several efforts have been made to improve the thermoelectric performance of $FeSb_2$ using the techniques of doping[5-8], nanoinclusions [9], nanostructuring [10, 11] and stoichiometric adjustment [12]. Studies have shown Te to be one of the most effective dopants to improve *ZT* of $FeSb_2$ by reducing the thermal conductivity via point defect scattering. Sun *et al*. [6] reported a *ZT* value of ~ 0.013 at around 100 K in $FeSb_{2-x}Te_x$ single crystal samples for their optimized doping concentration of x = 0.16. In our earlier work [10], we were able to reduce the thermal conductivity by three orders of magnitude to increase the peak value of *ZT* by 160 % (0.013 at 50 K) in nanostructured samples when compared to the single crystal counterpart. Recently [13], we also demonstrated the concept of semiconductor/metal interface in $FeSb_2$ to further enhance *ZT* by 70 % (0.02 at 50 K).

Mechanical nanostructuring has been very effective to reduce the thermal conductivity of $FeSb_2$ but not to improve the *ZT* value [10]. This is because as the grain size decreases, the Seebeck peaks are drastically suppressed and as a result, the gain in thermal conductivity reduction is negated by the loss in power factor ($S^2\rho^{-1}$). Based on our analysis[14], this is attributed to the fact that, the increased scattering of the phonons from the grain-boundary suppresses the phonon-drag contribution to the Seebeck coefficient. It seems like a significant further increase in *ZT* of $FeSb_2$ through nanostructuring alone is difficult. On the other hand, it has been demonstrated that Te-substituted single crystal of $FeSb_2$ exhibits a reduced thermal conductivity with a relatively large value of power factor[6, 7]. In this report, we present the combined effect of mechanical nanostructuring and Te-doping on the thermoelectric properties of $FeSb_2$. For this, we use the optimized Te-doped stoichiometric composition reported for single crystal in ref. [6] and tune the thermoelectric properties of nanostructured samples by changing the hot pressing temperature. Our results show a significant drop in thermal conductivity and an enhanced value of the power factor to further improve *ZT* values.

## Experimental

Stoichiometric amounts of Fe, Sb and Te were mixed and melted at 1000 °C inside an evacuated and sealed quartz tube. The tube is quenched in cooling water for rapid cooling and solidification. After solidification the ingot was ball milled for 15 hours. The resulting nanopowder was DC hot pressed under a pressure of 80 MP for 2 minutes at several different temperatures (200, 400, 500 and 600 °C) to vary the grain size in the nanocomposite sample disks. The samples were sputtered with gold to optimize electrical and thermal contacts and then cut into rectangular samples (2x2x8 mm³). The Seebeck Coefficient (*S*), electrical resistivity ($\rho$), and thermal conductivity ($\kappa$) were measured using thermal transport option (TTO) of the Physical Property Measurement System (PPMS). The horizontal rotator option was used to measure Hall coefficient ($R_H$) with typical dimensions of 1×2×10 mm³.

## Results and Discussion

Figure (1) shows the temperature dependence of the thermal conductivity for samples pressed at different temperatures. The thermal conductivity values of all the nanostructured samples are significantly reduced compared to the values reported for single crystal [4]. For sample $FeSb_2$ HP 500, the thermal conductivity at 100 K is lowered by 76 % from 30 W $m^{-1}$ $K^{-1}$ for single crystal to 7.08 W $m^{-1}$ $K^{-1}$. For the representative sample $FeSb_{1.84}Te_{0.16}$ HP 500, κ = 4.38 W $m^{-1}$ $K^{-1}$ at 100 K. The reduced thermal conductivity in our doped nanostructured samples is associated with two dominant scattering mechanisms: grain-boundary scattering and point-defect scattering. The reduction from 30 W $m^{-1}$ $K^{-1}$ to 7.08 W $m^{-1}$ $K^{-1}$ comes solely from increased scattering of the phonons off the grain boundaries introduced via nanostructuring. By Te-doping, the thermal conductivity of the nanostructured sample is further reduced from 7.08 W $m^{-1}$ $K^{-1}$ to 4.38 W $m^{-1}$ $K^{-1}$ at 100 K, a decrease by 38 %. A significant suppression of thermal conductivity in Te-doped $FeSb_2$ single crystals has been discussed in detail previously by Sun et al.[6] They attributed such a reduction to the introduced charge carriers rather than chemical disorder whereas Wang et al.[7] attributed this to the enhanced point defect scattering caused by a different bonding tendency and thermal conductivities of Sb and Te.

Figure 2 shows the temperature dependence of Seebeck coefficient. At 300 K, the Seebeck coefficient has a small positive value (p-type) ~2 μV $K^{-1}$ for all the Te-doped samples. As the temperature decreases, the Seebeck coefficient decreases and change to negative (n-type) value at ~290 K. The Seebeck coefficient assumes a peak value at 90 K for all the samples. The largest peak value for the Seebeck coefficient among our samples is ~ -107 μV $K^{-1}$ for sample $FeSb_{1.84}Te_{0.16}$ HP 600 which is two orders of magnitude less than the reported value for undoped $FeSb_2$ single crystals [4] and is one-fourth of the value (~ -400 μV$K^{-1}$) for $FeSb_{1.84}Te_{0.16}$ single crystals [6]. The peak value of the Seebeck coefficient decreases with decreasing hot pressing temperature. This decrease, based on our analysis, comes from two factors: increased carrier density due to decreased grain size (associated with defects) [10] and suppression of the phonon-drag contribution due to increased grain boundary scattering at smaller grain size level[14, 15]. We are aware that the role of phonon-drag effects in the thermoelectric properties of $FeSb_2$ single crystals was reported to be minor by many authors [16-18], however our data on nanostructured samples furnish evidence supporting the presence of phonon drag effects in $FeSb_2$ [14, 15]. The inset of Figure 2 shows the temperature dependent Seebeck coefficient for representative sample $FeSb_{1.84}Te_{0.16}$ HP 500 and its undoped counterpart $FeSb_2$ HP 500. The peak in the Seebeck coefficient becomes smaller, broader, and shifts to higher temperature upon Te-doping. Shifting of the Seebeck peak with increasing Te-content has also been reported in ref. [6]. With Te doping, the system evolves from semiconductor to metal with increased carrier density which causes decrease in Seebeck coefficient.

The temperature dependence of the electrical resistivity for the four doped nanostructured samples is shown in Figure 3. The electrical resistivity of the undoped sample FeSb$_2$ HP 500 is represented by the right y-axis. The undoped sample exhibits a semiconducting behavior throughout the temperature range 5-300 K with increasing resistivity as temperature decreases. A sharp increase in resistivity below 70 K indicates an insulating ground state. In contrast to the undoped FeSb$_2$, the Te doped samples exhibit suppressed electrical resistivity with a metallic ground state. As a result, at 5 K, the resistivity in doped sample drops by two orders of magnitude as compared to that of the undoped sample. At 100 K, however, the resistivity decreases by one order of magnitude only. The semiconductor to metal transition temperature for different samples falls within the range of 100 – 130 K. A slight shift in transition temperature to higher temperature occurs as the hot pressing temperature increases. Here we note that such a semiconductor to metal transition induced by Te-doping was reported earlier by Hu et al.[19]. Their extensive study on magnetic and electrical properties of Fe(Sb$_{1-x}$Te$_x$)$_2$ single crystal revealed a metallic ground state for $0.01 \leq x \leq 0.2$ below 200 K. The Te-concentration in our sample (x = 0.08) lies within the above range and is consistent with the results in ref. [16]. The inset of Fig. 3 shows the fitting to the Arrhenius law, $\rho = \rho_o \exp[-E_g/2k_BT]$, for the undoped sample FeSb$_2$ HP 500. Here, $E_g$ is the activation energy gap, $\rho_o$ is the residual resistivity at absolute zero and $k_B$ = 1.38 ×10$^{-23}$ JK$^{-1}$ is the Boltzmann constant. A small energy gap of 0.6 meV(smaller than 4-6 meV reported for single crystal FeSb$_2$) is observed at low temperature. Such a small value for the energy gap favors the semiconductor to metal transition as explained by the authors in Ref. [6]. At higher temperature, $E_g$ ~ 23 meV, which lies within the value of 20-40 meV reported for FeSb$_2$ single crystal.

The carrier concentration ($n$) and the Hall mobility ($\mu$) were estimated from the Hall coefficient ($R_H$) and resistivity ($\rho$) measurements using $n = 1/|R_H|e$ and $\mu = |R_H|/\rho$, under the single parabolic band model. Here $e$ = 1.6 ×10$^{-19}$ C is the electronic charge. Figure 4 shows the temperature dependence of $n$ and $\mu$ for the undoped and doped nanostructured samples pressed at the same temperature of 500 °C, in the selected temperature range of 60 – 200 K. As expected, the carrier concentration of the doped sample is increased by one order of magnitude around 100 K, the temperature of interest. Also, the carrier concentration is less temperature sensitive for doped samples, a result consistent with the metallic nature as seen in electrical resistivity data at lower temperature. The Hall mobility is reduced in the doped sample, at 100 K, $\mu$ = 5.3 cm$^2$ V S$^{-1}$ and 3.3 cm$^2$ V S$^{-1}$ for samples FeSb$_2$ HP 500 and FeSb$_{1.84}$Te$_{0.16}$ HP 500, respectively. In general, the Seebeck coefficient decreases with increase in carrier concentration and decreases with decrease in carrier mobility. Therefore the suppressed peak values for the Seebeck coefficient in doped sample is due to high carrier concentration.

Figure 5 shows the temperature dependence of the power factor ($S^2\rho^{-1}$). When compared to the undoped nanostructured samples, the power factors in Te-doped samples increased significantly. For example, $S^2\rho^{-1}$ = 9.9 × 10$^{-4}$ W m$^{-2}$ K$^{-1}$ at 80 K for the representative sample FeSb$_{1.84}$Te$_{0.16}$ HP 500, which is an increase of 386 % from the corresponding value for the undoped sample FeSb$_2$ HP 500. Although, the peak value of the Seebeck coefficient decreases with Te-doping, such an enhanced power factor comes from the reduced electrical resistivity in the doped samples. Among the Te-doped samples, the peak values for the power factor decreases upon decreasing the grain size and is consistent with the trend seen in electrical resistivity and Seebeck coefficient data.

We learned from our previous work[10] that thermal conductivity and the Seebeck coefficient in $Z = S^2\rho^{-1}\kappa^{-1}$ contribute dynamically as a function of grain size in FeSb$_2$. Thermal conductivity can be reduced dramatically through grain size reduction but simultaneously the power factor decreases drastically. Therefore, a better compromise between these dynamic properties can be anticipated somewhere in the optimized grain size level, where moderate values for both the thermal conductivity and power factor can be maintained. In Figure 6, we have presented *ZT* as a function of temperature. For all the doped samples, the curve assumes a peak value (*ZT*$_{max}$) at around 100 K. For the optimized sample FeSb$_{1.84}$Te$_{0.16}$ HP 500, *ZT*$_{max}$ = 0.022 at 100 K. This is an increase of 62 % over the optimized value for the Te-doped single crystal of 0.012 in ref. [6]. When compared to *ZT* of FeSb$_2$ HP 500 (0.0017 at 25 K), the *ZT* values for the optimized sample FeSb$_{1.84}$Te$_{0.16}$ HP 500 is improved by 11 times. Such a significant improvement was achieved by both reducing thermal conductivity and increasing the power factor.

## Conclusion

We studied the thermoelectric transport properties of FeSb$_{1.84}$Te$_{0.16}$ nanostructured samples prepared by high temperature alloying, ball milling and hot pressing at different temperatures. Te-doping induced semiconductor to metal transition caused an increased electrical conductivity but decreased Seebeck coefficient. Our results revealed decreased Hall mobility and increased Hall carrier density in doped samples. Overall, the power factor values were improved in doped samples. On the other hand, already reduced thermal conductivity of nanostructured samples was further reduced upon Te-doping. As a result we were able to increase *ZT* to 0.022 at 100 K for the optimized sample FeSb$_{1.84}$Te$_{0.16}$ HP 500 which is an increase by 62 % over the single crystal counterpart.

**Acknowledgment**

We gratefully acknowledge funding for this work by the Department of Defense, United States Air Force Office of Scientific Research's MURI program under contract FA9550-10-1-0533.

# Figure Captions

Figure 1. Thermal conductivity as a function of temperature for the four $FeSb_{1.84}Te_{0.16}$ samples hot pressed at different temperatures. Data for the $FeSb_2$ sample hot pressed at 500 °C is also shown for comparison.

Figure 2. Seebeck coefficient as a function of temperature for the four $FeSb_{1.84}Te_{0.16}$ samples. The inset shows the comparison between the temperature dependence of the representative sample $FeSb_{1.84}Te_{0.16}$ HP 500 and an undoped counterpart $FeSb_2$ HP 500.

Figure 3. Electrical resistivity (left y-axis) as a function of temperature for the four $FeSb_{1.84}Te_{0.16}$ samples. The right y-axis corresponds to the electrical resistivity for the sample $FeSb_2$ HP 500. The inset shows the Arrhenius plot for the sample $FeSb_2$ HP 500.

Figure 4. Carrier concentration (left y-axis) and Hall mobility (right y-axis) as a function of temperature in the temperature range of 60-200 K.

Figure 5. Power factor as a function of temperature for the four $FeSb_{1.84}Te_{0.16}$ samples. Data for $FeSb_2$ HP 500 is also included for comparison.

Figure 6. *ZT* as a function of temperature for the four $FeSb_{1.84}Te_{0.16}$ samples. Data for $FeSb_2$ HP 500 and $FeSb_{1.84}Te_{0.16}$ single crystal (taken from ref. [6]) are also included for comparison.

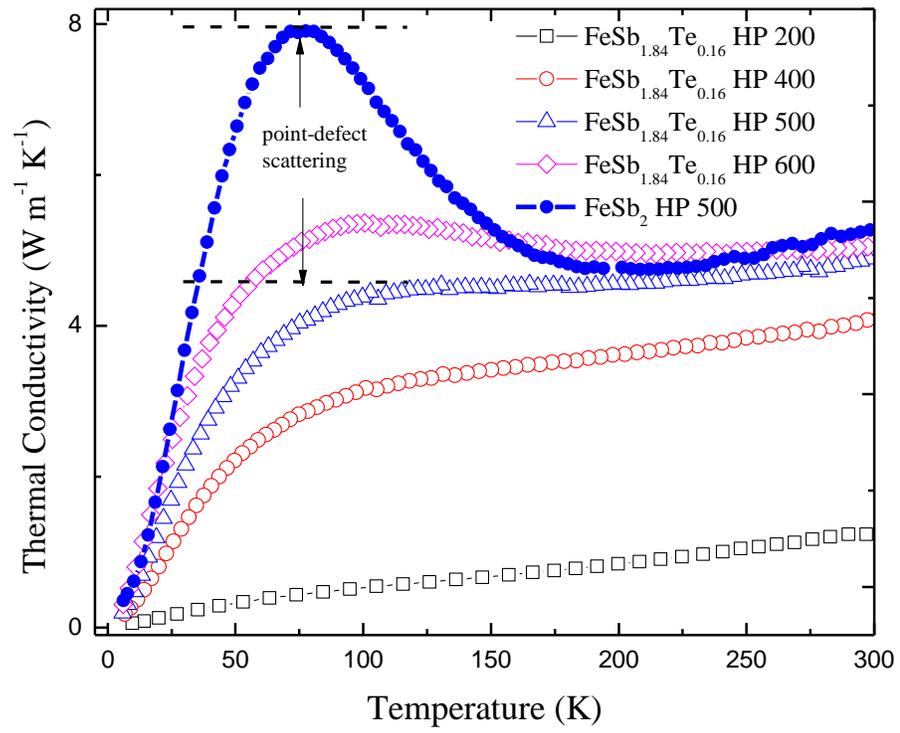

Figure 1

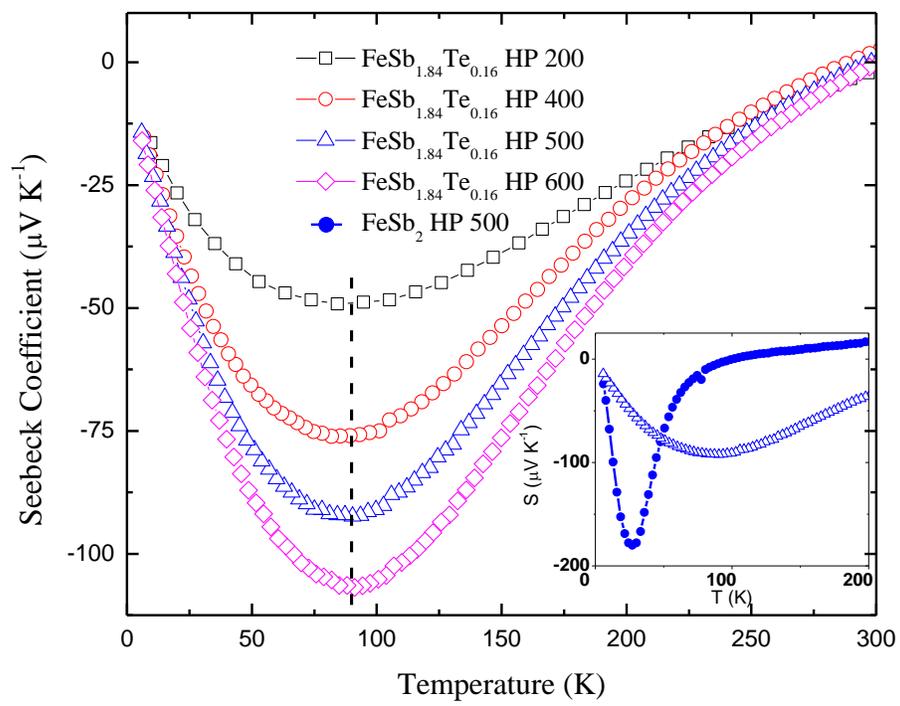

Figure 2

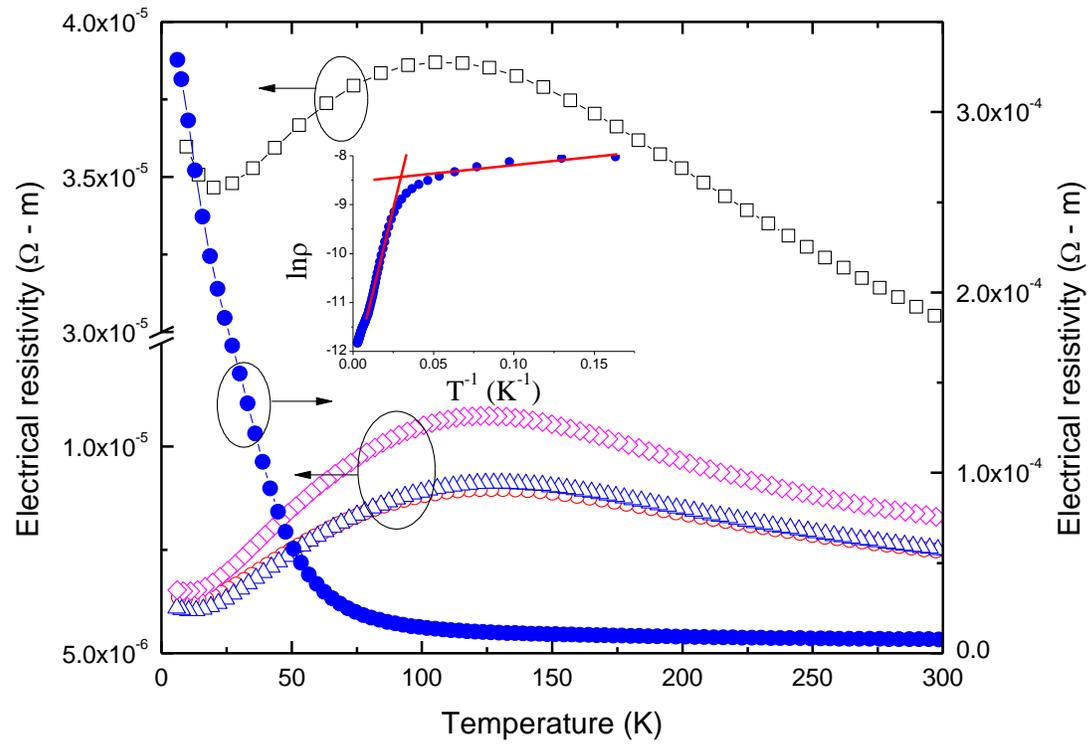

Figure 3

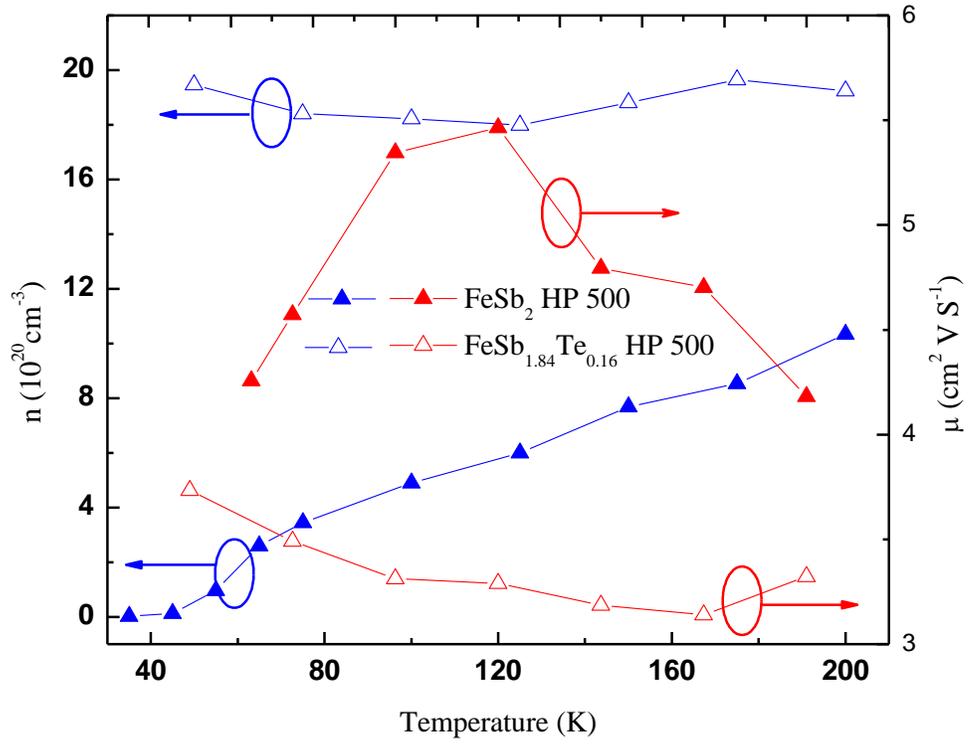

Figure 4

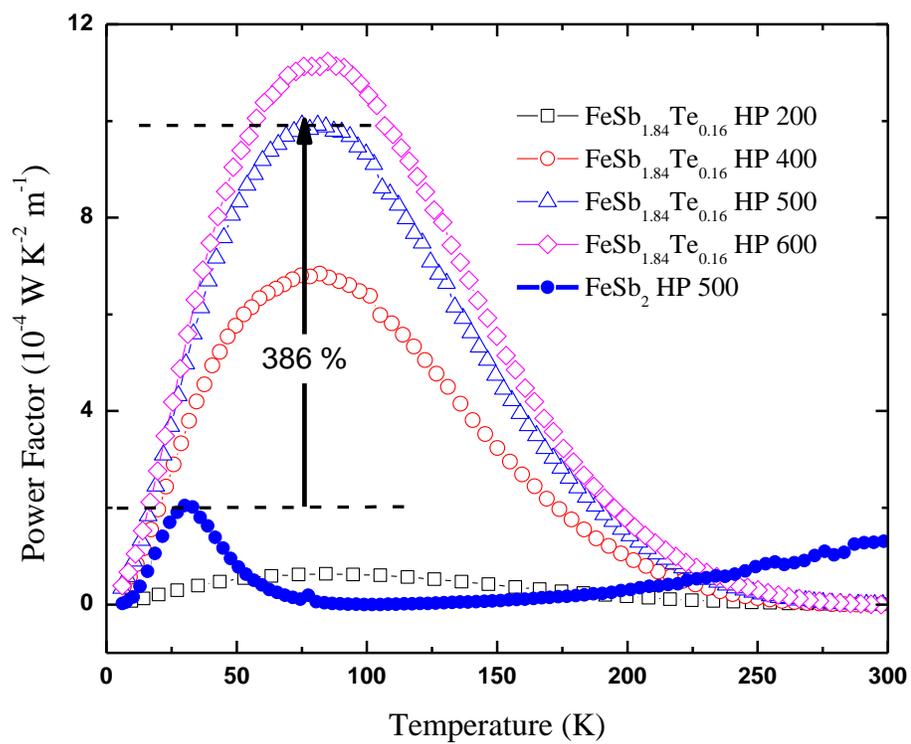

Figure 5

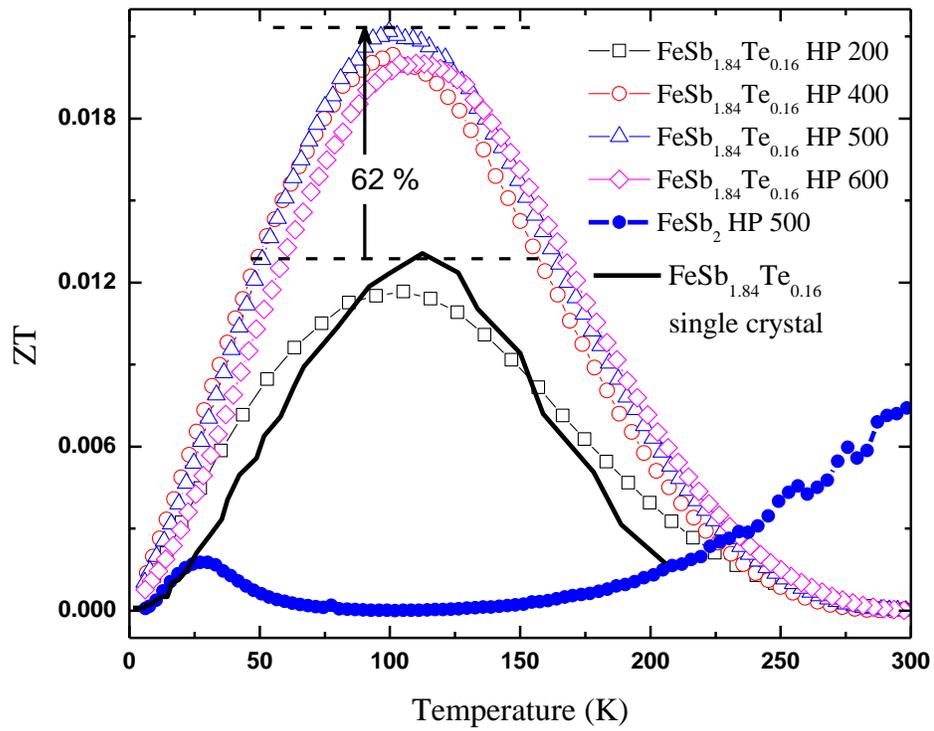

Figure 6